\documentclass[ showpacs, preprintnumbers, amsmath, amssymb,twocolumn]{revtex4}
\usepackage{amssymb}

\usepackage{graphicx}
\usepackage{dcolumn}
\usepackage{bm}

\begin{document}

\preprint{...}

\title{Exotic quantum statistics of composite particles and frustrated quasiparticles}

\author{Tieyan Si}
\affiliation{Academy of Fundamental and Interdisciplinary Sciences,
Harbin Institute of Technology, Harbin, 150080, China}

\date{\today}

\begin{abstract}

We study the exotic quantum statistical behavior of composite
particle of double-spin cluster and quasiparticle of triple-spin
cluster in a four-spin quantum model. We constructed a four spin-1/2
model on a triangular star lattice but added frustrated coupling
terms of plaquette quasiparticles. The eigenstates of this model are
maximal entangled quantum states like Greenberger-Horne-Zeilinger
state and Yeo-Chua's genuine four-qubit entangled state. We
generalized the conventional definition for quantum statistics of
two elementary particles to composite particle of multispin
clusters. Greenberger-Horne-Zeilinger state and Yeo-Chua's genuine
four-qubit entangled state showed different behavior according to
this generalized definition. The quantum statistical behavior of the
composite particle of double-spin cluster is neither boson nor
fermion in ground state and some intermediate excited states. The
triple-spin cluster of this model is eigen-quasiparticles. We
perform permutation operation on the eigenstates of triple-spin
plaquette operator according to this generalized definition for
quantum statistics of multi-spin clusters, the statistical matrix of
exchanging two triple-spin quasiparticles is far beyond fermion and
boson. The von Neumann entropy of the triple-spin quasiparticle is
also highly nontrivial. These nontrivial quantum statistical
behavior of plaquette quasiparticles is helpful for decoding the
non-abelian anyons in Kitaev honeycomb model.

\end{abstract}


\maketitle

\section{Introduction}

Fermion is the most fundamental building block for matters. Most
bosons are composite particle of even number of fermions. The
typical fermion, proton, is also a composite particle of three
quarks. The collective wave function of two fermions is
anti-symmetric, $\psi(r_{1},r_{2})=-\psi(r_{2},r_{1})$. While the
collective wave function of two bosons is symmetric,
$\psi(r_{1},r_{2})=\psi(r_{2},r_{1})$. The statistical factor of
anyon is defined as
$\psi(r_{1},r_{2})=e^{i\theta}\psi(r_{2},r_{1})$, where
[$\theta=\alpha\pi$,$1<\alpha<2$]. Different statistical behavior of
these composite particles leads to different physical phenomena. The
unknown statistical behavior of composite particles gained
long-lasting research interest until today. One recent study defines
an arbitrary composite boson by the product of two fermion or boson
operator \cite{sylee}, interesting eigenstate and commutator was
derived by defining an effective composite boson annihilation
operator \cite{sylee}. If Pauli principle has no influence on
physical behavior of many composite bosons, the composite boson of
two entangled fermions can be treated as an elementary bosonic
particle \cite{Tichy}. The upper and lower bounds of a defined
quantity determines the bosonic statistical behavior of a pair of
entangled fermions \cite{Chudzicki}. Many composite particles in
strongly correlated quantum many body system exhibits exotic
statistical behavior beyond boson and fermion, such as quasiparticle
and quasiholes in fractional quantum Hall system \cite{Laughlin},
abelian anyon model for topological quantum computation
\cite{Nayak}, and so on. The non-abelian plaquette excitations in
Kitaev's honeycomb model obeys non-trivial topological fusion rule.
However it is hard to find the exact relationship between the
topological fusion rule and the conventional textbook definition of
anyons above. In this paper, we proposed a generalized definition of
the textbook's definition of anyon for studying the composite
particle of double-spins and triple-spin quasiparticles in a
triangular star model with frustrated quasiparticle, which can be
viewed as the minimal model of Kitaev honeycomb model \cite{kitaev},
but the additionally introduced frustration terms makes it more
complicate than Kitaev honeycomb model.

It is the quantum entanglement between the elementary particles of a
composite particle that drives the statistical behavior of composite
particle out of the scope of boson and fermion. Quantum entanglement
attracted many research interests for quantum computation
\cite{Horodecki}\cite{Nest}. The upper bounds of squared concurrence
\cite{CJZhang} and its dual lower bound of squared concurrence for
arbitrary mixed state \cite{Mintert} can be used to estimate the
entanglement in experimental measurement. Modern quantum optical
technology can implement finite number of qubits \cite{pan}.
Greenberger-Horne-Zeilinger state has been generated in laboratory
\cite{GHZstate}, so did the W-state of four qubits \cite{Wstate}.
The Cross-Kerr nonlinearity of quantum optics \cite{zhao} is
suggested for implementing Yeo-Chua's genuine four-qubit entangled
states \cite{yeo}. Since the four-qubit states is very convenient
for experiment implementation, theoretical research interest on
quantum entanglement of four-qubit state is accumulating rapidly
\cite{Horodecki}.

Recently the classification of quantum entangled states by symmetry
has aroused many research interest \cite{Vladislav}
\cite{Aulbach}\cite{Migda}\cite{Jeong}\cite{wang}\cite{Markham}.
Entangled state is categorized by the permutation symmetry of the
subsystems \cite{Aulbach}. the Greenberger-Horne-Zeilinger-Like
symmetric state of four qubits is constructed By extending the
symmetry of three-qubit Greenberger-Horne-Zeilinger (GHZ) state to
four qubits \cite{Jeong}. While most of these states is constructed
without Hamiltonian. In this triangular model, we actually derived
many entangled states by directly solving this four-spin quantum
model. We generalize the textbook definition of permutating two
particles in collective wave function to permutating two pairs of
particles, and also generalize it to permutating two composite
quasiparticle of three-spin cluster. With the help of Hamiltonian
operator and plaquette operator, we found the generalized triplet
state and singlet state of the composite spin of double-spins.

This four-spin model is inspired by Kitaev honeycomb lattice model
for topological quantum computation \cite{kitaev}\cite{Nayak}. We
built the coupled four spin model on a triangular star lattice. The
first part of the Hamiltonian obeys the coupling rule of Kitaev
honeycomb model \cite{kitaev}. The newly added second part is
anti-ferromagnetic coupling between neighboring plaquette operators
which sits right at the center of plaquette. The anti-ferromagnetic
coupling between particles on triangular lattice results in
geometrically frustrated quantum system \cite{Moessner}. The
geometric frustration between plaquette quasiparticles of this
four-spin model also introduced interesting statistics of
quasiparticles beyond that of Kitaev honeycomb model, for instance,
the ground state is no longer homogeneous gauge pattern.

The article is organized as following: In section II, we proposed
the triangular star model and the generalized definition for
statistics of composite particle of double-spins. The nontrivial
statistical matrix of composite particle of double-spins was
computed. In section III, we computed the quantum statistical matrix
of the plaquette quasiparticle of triple-spin clusters. A
non-trivial statistical matrix and von Neumann entropy was found.
Section IV is a brief summary.

\section{The quantum statistics of composite double-spin operators in a triangular star model}

\begin{figure}[htbp]
$
\begin{array}{c@{\hspace{0.15in}}c}
\includegraphics[scale=0.23]{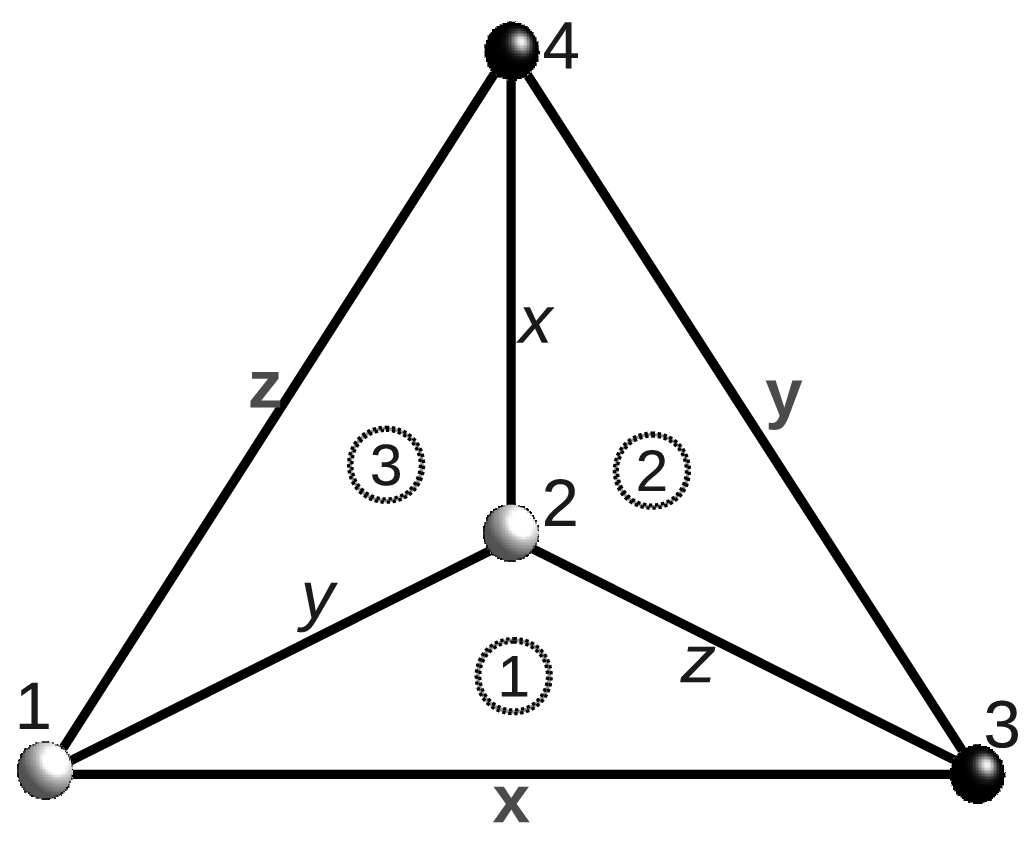} &
\includegraphics[scale=0.20]{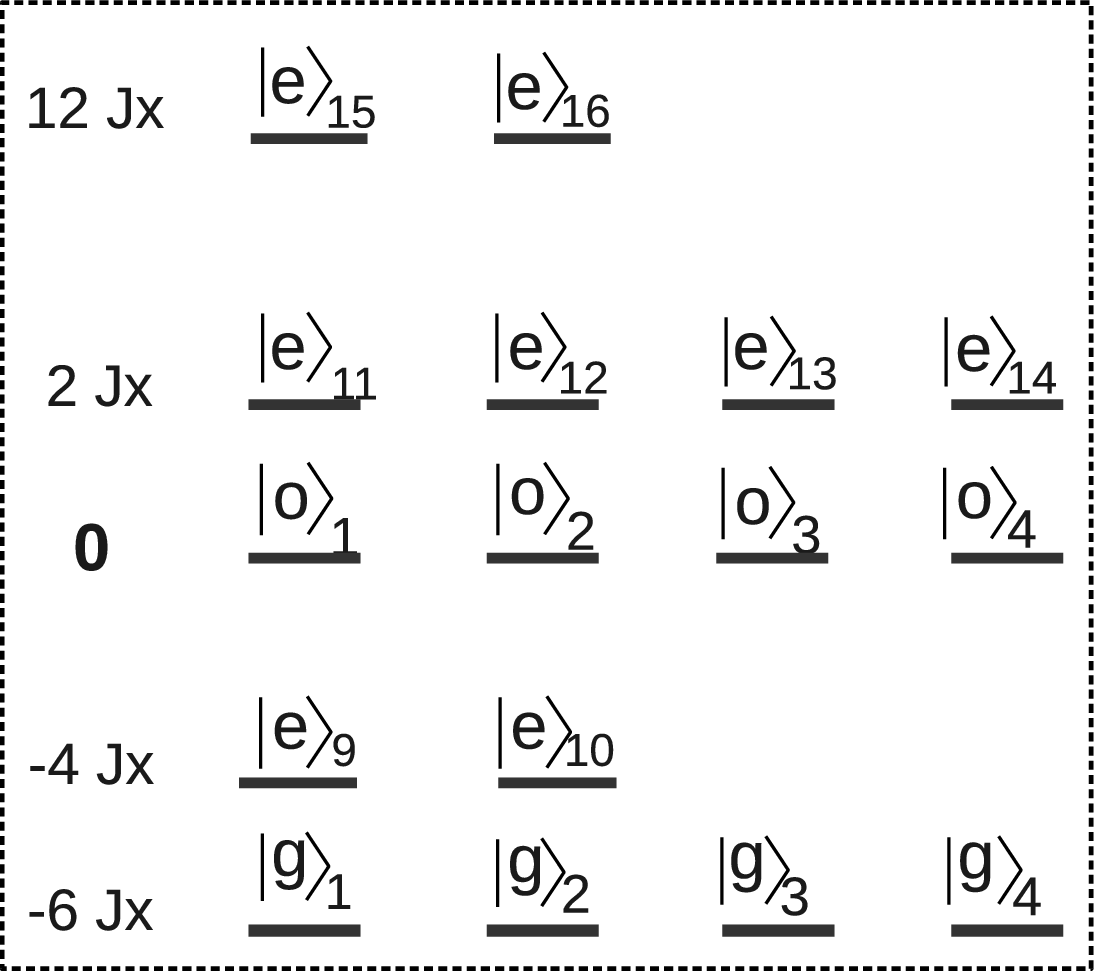}\\
\mbox{\bf (a)} & \mbox{\bf (b)}
\end{array}
$ \caption{\label{four}(a) The triangular star model. Quasiparticle
sits right at the center of each triangle. (b) The eigenenergy
levels of the triangular star model with respect to the parameter
setting of $(J_{p}=2 J_{x},\; J_{z}=2 J_{x},\;J_{y}=2 J_{x})$.}
\end{figure}

This triangular star model places the four particles at the vertices
of a triangular star lattice(Fig. \ref{four}). The triangular star
has three independent triangular plaquette. The four particles
coupled to each other following Kitaev honeycomb model. We added an
antiferromagnetic coupling between the nearest neighboring plaquette
on the Hamiltonian,
\begin{eqnarray}\label{Htri}
H_{a}&=&J_{x}\sigma^{x}_{1}\sigma^{x}_{3}
+J_{y}\sigma^{y}_{1}\sigma^{y}_{2}
+J_{z}\sigma^{z}_{2}\sigma^{z}_{3}
+J_{x}\sigma^{x}_{2}\sigma^{x}_{4}
+J_{y}\sigma^{y}_{3}\sigma^{y}_{4}\nonumber\\
&+&J_{z}\sigma^{z}_{1}\sigma^{z}_{4}
+J_{p}\hat{S}_{\textbf{1}}\hat{S}_{\textbf{2}}
+J_{p}\hat{S}_{\textbf{2}}\hat{S}_{\textbf{3}}
+J_{p}\hat{S}_{\textbf{3}}\hat{S}_{\textbf{1}}.
\end{eqnarray}
The three plaquette operators $\hat{S}_{\textbf{j}}$ are quantum
string operators around each triangular plaquette,
\begin{equation}\label{3s}
\hat{S}_{\textbf{1}}=\sigma^{z}_{1}\sigma^{x}_{2}\sigma^{y}_{3},\;\;
\hat{S}_{\textbf{2}}=\sigma^{z}_{4}\sigma^{y}_{2}\sigma^{x}_{3},\;\;
\hat{S}_{\textbf{3}}=\sigma^{x}_{1}\sigma^{z}_{2}\sigma^{y}_{4}.\;\;
\end{equation}
They commute with Hamiltonian and commute with each other, i.e.,
$[\hat{S}_{\textbf{i}},H]=0$,
$[\hat{S}_{\textbf{i}},\hat{S}_{\textbf{j}}]=0,\;i\neq{j}=1,2,3$.

The three conserved plaquette operator divide the total Hilbert
space into three sectors. Each plaquette operator has eigenvalues
$+1$ and $-1$ within its Hilbert space,
$\hat{S}_{\alpha}|\psi\rangle=\pm 1|\psi\rangle$. Every triangular
plaquette operator defines an effective Ising spin within each
sector. In the Kitaev honeycomb model, the ground state chooses a
homogeneous gauge pattern. All plaquette operators take the same
eigenvalue. As there is antiferromagnetic coupling between two
plaquette, the ground state is no longer the homogeneous gauge
pattern. The ground state bear the frustrated gauge pattern.

We first solve the model by diagonalizing the Hamiltonian matrix.
The spin operators take a sixteen dimensional representation,
\begin{eqnarray}
  \sigma^{\mu}_{1} &=& \hat{\sigma}^{\mu}_{1}\otimes{\textbf{I}_{2}}
   \otimes{\textbf{I}_{3}}\otimes{\textbf{I}_{4}},\;\;\;
  \sigma^{\mu}_{2} = {\textbf{I}_{1}}\otimes\hat{\sigma}^{\mu}_{2}
   \otimes{\textbf{I}_{3}}\otimes{\textbf{I}_{4}},\nonumber\\
  \sigma^{\mu}_{3} &=& {\textbf{I}_{1}}\otimes{\textbf{I}_{2}}
   \otimes\hat{\sigma}^{\mu}_{3}\otimes{\textbf{I}_{4}}, \;\;\;
  \sigma^{\mu}_{4} = {\textbf{I}_{1}}\otimes{\textbf{I}_{2}}
   \otimes{\textbf{I}_{3}}\otimes\hat{\sigma}^{\mu}_{4},\;\;\;
\end{eqnarray}
where $\hat{\sigma}^{\mu}_{i}$ are the conventional Pauli matrices
and ${\textbf{I}_{i}}$ is the 2$\times$2 identity matrix. The symbol
$\otimes$ denotes direct product. The eigenvalues of the sixteen
dimensional Hamiltonian matrix lead to eight discrete energy levels,
\begin{eqnarray}
E^{\pm}_{p}&=&3J_{p}\pm2\sqrt{J^{2}_{x}+J^{2}_{y}+J^{2}_{z}},\;\;\nonumber\\
E^{\pm}_{\mu}&=&-J_{p}\pm 2 J_{\mu},\;\;\;\;\;\mu=x,y,z.
\end{eqnarray}
Each energy level has two fold degeneracy. The energy levels are
listed in Fig. \ref{four} (b). The eigenenergy and eigenstates are
computed directly from the Hamiltonian matrix. The newly added
coupling terms of two plaquette operators commute with Hamiltonian.
It does not modify the physics mechanism of Kitaev honeycomb model.
The vortex excitation in the triangular plaquette represents the
same type of quasiparticle excitation of Kitaev honeycomb model. The
main different character of this triangle star model from Kitaev
honeycomb model is the neighboring quasiparticles are now
antiferromagnetically coupled to each other. As the three plaquette
are placed on a triangle, it forms a typical pattern of frustrated
Ising spin system. If the three quasiparticles have the same
eigenvalue, it is a fully frustrated system since the spins intend
to be parallel to each other. If there is only one pair of spin are
oriented at the same direction, it is the minimally frustrated
state. The ground state is the minimally frustrated quasiparticle
state for which the system bears minimal energy.

The two levels with higher eigenenergy $E^{\pm}_{p}$ correspond to
excited states for which the eigenvalues of the three plaquette
operators takes the same value
${S}_{\textbf{1}}={S}_{\textbf{2}}={S}_{\textbf{3}}=\pm1$. The three
pairs of energy level with lower energy $E^{\pm}_{\mu}$ correspond
to the frustrated gauge pattern. Only one pair of quasiparticles is
not frustrated for $E^{\pm}_{p}$. If $J_{x}=J_{y}=J_{z}$, the six
energy levels of $E^{\pm}_{\mu}$ would be degenerated.

If the quasiparticle coupling interaction $J_{p}$ becomes zero, the
triangular star model reduces to a finite Kitaev honeycomb model on
triangular star lattice. For a stronger quasiparticle coupling
interaction than the spin coupling,
\begin{equation} \label{jp=2x}
J_{p}=2 J_{x},  \;\;\; J_{z}=2 J_{x}, \;\;\;J_{y}=2 J_{x},
\end{equation}
there exists four degenerated states with zero energy. Without
losing important physics, we focus on this special parameter setting
of Eq. (\ref{jp=2x}) in the following. By representing the
eigenvectors of the 16 dimensional Hamiltonian matrix by the 16
four-spin basis, we derived the spin configurations corresponding to
the four zero energy states,
\begin{eqnarray}
|o\rangle_1&=&\frac{1}{2\sqrt{15}}[|\Uparrow\Uparrow\rangle+|\Downarrow\Downarrow\rangle
+2|\circlearrowright\circlearrowleft\rangle+2|\circlearrowleft\circlearrowright\rangle
   \nonumber\\
&-&5|\circlearrowleft\circlearrowleft\rangle-5|\circlearrowright\circlearrowright\rangle], \nonumber\\
|o\rangle_2&=&\frac{1}{2}[|\Uparrow\circlearrowright\rangle+|\circlearrowright\Uparrow\rangle
+|\circlearrowleft\Downarrow\rangle+|\Downarrow\circlearrowleft\rangle], \nonumber\\
|o\rangle_3&=&\frac{1}{2}[|\Uparrow\circlearrowleft\rangle+|\circlearrowright\Downarrow\rangle+
|\circlearrowleft\Uparrow\rangle+|\Downarrow\circlearrowright\rangle], \nonumber\\
|o\rangle_4&=&\frac{1}{2\sqrt{3}}[|\Uparrow\Downarrow\rangle+|\Downarrow\Uparrow\rangle
+2|\circlearrowleft\circlearrowleft\rangle+2|\circlearrowright\circlearrowright\rangle \nonumber\\
&-&|\circlearrowright\circlearrowleft\rangle-|\circlearrowleft\circlearrowright\rangle]
.
\end{eqnarray}
Here the symbols in the quantum wave function represent the double
spin configurations,
\begin{eqnarray}
|\Uparrow\rangle=|\uparrow\uparrow\rangle,\;|\Downarrow\rangle=|\downarrow\downarrow\rangle,\;
|\circlearrowright\rangle=|\uparrow\downarrow\rangle,\;|\circlearrowleft\rangle=|\downarrow\uparrow\rangle.
\end{eqnarray}
The eigenenergy of the states above is zero, i.e.,
$H|o\rangle_\mu=0,\;\;\mu=1,2,3,4.$ The zero energy state pair of
$|o\rangle_2$ and $|o\rangle_3$ comes from the minimal frustrated
gauge pattern, i.e, $E^{+}_{x}=-J_{p}+ 2 J_{x}=0$. The two zero
energy states of $|o\rangle_1$ and $|o\rangle_4$ correspond to the
fully frustrated states $E^{-}_{p}=3J_{p}-6J_{x}=0$.

The ground state has four fold degeneracy. We denote them as
$|g\rangle_\mu,\;\;\mu=1,2,3,4.$,
\begin{eqnarray}
&&|g\rangle_1=\frac{1}{2}[|\Downarrow\circlearrowleft\rangle-|\Uparrow\circlearrowright\rangle+|\Uparrow\circlearrowleft\rangle
-|\Downarrow\circlearrowright\rangle],\nonumber\\
&&|g\rangle_2=\frac{1}{\sqrt{2}}[|\Downarrow\Uparrow\rangle-|\Uparrow\Downarrow\rangle], \nonumber\\
&&|g\rangle_3=\frac{1}{{2}}[|\circlearrowleft\Uparrow\rangle-|\circlearrowright\Downarrow\rangle
+|\circlearrowleft\Downarrow\rangle-|\circlearrowright\Uparrow\rangle], \nonumber\\
&&|g\rangle_4=\frac{1}{\sqrt{2}}[|\circlearrowleft\circlearrowleft\rangle
-|\circlearrowright\circlearrowright\rangle].
\end{eqnarray}

The eigenenergy of the four ground states are $E_g=-6J_x$, i.e.,
$\;H|g\rangle_\mu=-6J_x|g\rangle_\mu,\;\;\mu=1,2,3,4.$ The ground
states are the minimally frustrated quasiparticle states. Two of
them comes from $E^{-}_{y}=-J_{p} - 2 J_{y}=-6J_x$. The other two
states is for $E^{-}_{z}=-J_{p} - 2 J_{z}=-6J_x$. The spin
configuration of the two states $|g\rangle_1$ and $|g\rangle_3$ is
the superposition of three-up-one-down state and three-down-one-up
state. The two states of $|g\rangle_2$ and $|g\rangle_4$ is the
superposition state of four spin basis with two-up spins and
two-down spins. If we flip the spins of all of the four ground
states, it generates a minus sign upon the original wave function.
So the ground state breaks $Z_{2}$ symmetry.

The whole Hilbert space of eigenstates can be classified into two
classes. The first class breaks the $Z_{2}$ symmetry. This class
only includes the ground state and the four degenerated states with
respect to $2J_x$. The other class keeps $Z_{2}$ symmetry. This
class includes the zero energy states, the highest excited states
and all the rest excited states.

The energy level $\omega=12J_x$ is the highest energy level
corresponding to the fully frustrated quasiparticle state. The spin
configuration corresponding to this two-fold degenerated level is
$|e\rangle_{15}$ and $|e\rangle_{16}$,
\begin{eqnarray}
|e\rangle_{15}&=&\frac{1}{2\sqrt{3}}[|\Uparrow\Uparrow\rangle+|\Downarrow\Downarrow\rangle
+2|\circlearrowright\circlearrowleft\rangle+2|\circlearrowleft\circlearrowright\rangle   \nonumber\\
&+&|\circlearrowleft\circlearrowleft\rangle+|\circlearrowright\circlearrowright\rangle], \nonumber\\
|e\rangle_{16}&=&\frac{1}{2\sqrt{15}}[|\Uparrow\Downarrow\rangle+|\Downarrow\Uparrow\rangle
+2|\circlearrowright\circlearrowright\rangle+2|\circlearrowleft\circlearrowleft\rangle
   \nonumber\\
&+&5|\circlearrowleft\circlearrowright\rangle+5|\circlearrowright\circlearrowleft\rangle],
\end{eqnarray}

The eigenstates of this model can be viewed as entangled quantum
states of two double-spin clusters. For example, the nearest excited
state above the zero energy state has four-fold degeneracy. The spin
configuration of the four states are
\begin{eqnarray}
|e\rangle_{11}&=&\frac{1}{\sqrt{2}}[-|\Uparrow\Uparrow\rangle+|\Downarrow\Downarrow\rangle], \nonumber\\
|e\rangle_{12}&=&\frac{1}{2}[-|\Uparrow\circlearrowright\rangle-|\Uparrow\circlearrowleft\rangle
+|\Downarrow\circlearrowright\rangle+|\Downarrow\circlearrowleft\rangle],\nonumber\\
|e\rangle_{13}&=&\frac{1}{2}[-|\circlearrowright\Uparrow\rangle+|\circlearrowright\Downarrow\rangle
-|\circlearrowleft\Uparrow\rangle+|\circlearrowleft\Downarrow\rangle],\nonumber\\
|e\rangle_{14}&=&\frac{1}{\sqrt{2}}[-|\circlearrowright\circlearrowleft\rangle+|\circlearrowleft\circlearrowright\rangle].
\end{eqnarray}
The corresponding eigenenergy with respect to  $\{|e\rangle_{11}$,
$|e\rangle_{12}$, $|e\rangle_{13}$, $|e\rangle_{14}\}$ is $E=2J_x$.
The spin configurations of $|e\rangle_{12}$ and $|e\rangle_{13}$ is
the superposition of three-up-one-down and three-down-one-up. The
state $|e\rangle_{11}$ can be viewed as a dual state of the
well-known Greenberger-Horne- Zeilinger state for four qubit
\cite{GHZstate},
\begin{eqnarray}
|GHZ\rangle=|\Uparrow\Uparrow\rangle+|\Downarrow\Downarrow\rangle.
\end{eqnarray}
$|e\rangle_{14}$ is a singlet pair state of two double-spin clusters
. The other two states, $|e\rangle_{12}$ and $|e\rangle_{13}$, can
be viewed as generalized W-state of four qubits \cite{Wstate},
\begin{eqnarray}
|W\rangle=(|\Uparrow\circlearrowright\rangle+|\Uparrow\circlearrowleft\rangle
+|\circlearrowright\Uparrow\rangle+|\circlearrowleft\Uparrow\rangle).
\end{eqnarray}

The two degenerated states of energy level $E=-4J_x$ are
$|e\rangle_{9}\}$ and $\{|e\rangle_{10}$, they both bear similar
state structure as W-state,
\begin{eqnarray}
|e\rangle_9&=&\frac{1}{{2}}[|\Uparrow\circlearrowright\rangle-|\circlearrowright\Uparrow\rangle
-|\circlearrowleft\Downarrow\rangle+|\Downarrow\circlearrowleft\rangle],\nonumber\\
|e\rangle_{10}&=&\frac{1}{{2}}[|\Uparrow\circlearrowleft\rangle-|\circlearrowleft\Uparrow\rangle-|\circlearrowright\Downarrow\rangle
+|\Downarrow\circlearrowright\rangle].
\end{eqnarray}
These eigenstates are genuine entangled states of four spins. The
zero energy states and highest excited states, $\{|o\rangle_1$,
$|o\rangle_4\}$ and $\{|e\rangle_{15}$, $|e\rangle_{16}\}$, can be
classified into the same class as Yeo-Chua's genuine four-qubit
entangled state \cite{yeo},
\begin{eqnarray}
|\chi^{00}\rangle&=&|\Downarrow\Downarrow\rangle-|\Downarrow\Uparrow\rangle-
|\circlearrowleft\circlearrowleft\rangle+|\circlearrowleft\circlearrowright\rangle\nonumber \\
&+&|\circlearrowright\circlearrowleft\rangle+|\circlearrowright\circlearrowright\rangle
+|\Uparrow\Downarrow\rangle+|\Uparrow\Uparrow\rangle.
\end{eqnarray}

The eigenstates of this model can be implemented by the same
operation as that for generating Yeo-Chua's genuine four-qubit
entangled state using cross-Kerr nonlinearity \cite{zhao}. These
states can be mapped into graph states following the strategy of
Ref. \cite{ye}, the observable operators in graph state theory has a
physical meaning in this quantum model. Usually Yeo-Chua's genuine
four-qubit entangled state or W-state is constructed without a
Hamiltonian, but here we can derive the eigenenergy of these states.
One can decompose an arbitrary entangled state as the superposition
of these eigenstates. The weight of each eigenstate is marked by its
eigenvalue. This offers us a new angle to see the internal structure
of the quantum entanglement.

The Wootters's concurrence provide a convenient way to quantify
quantum entanglement of two-qubit states \cite{Wootters},
\begin{eqnarray}
\tau_{abcd}=|\langle\psi|\hat{\sigma}^{y}_{a}\otimes{\hat{\sigma}^{y}_{b}}
   \otimes{\hat{\sigma}^{y}_{c}}\otimes{\hat{\sigma}^{y}_{d}}|\psi^{\ast}\rangle|^{2}.
\end{eqnarray}
This concurrence has a physical interpretation in this quantum model
since none of the eignestates here includes complex numbers. The
concurrence operator corresponds to conserved plaquette operator in
this triangular star model. The product of any two plaquette
operators is a string of four identical spin operators,
\begin{eqnarray}
\hat{S}_{\textbf{1}}\hat{S}_{\textbf{2}}&=&
\sigma^{z}_{1}\sigma^{z}_{2} \sigma^{z}_{3}\sigma^{z}_{4},\;\
\hat{S}_{\textbf{2}}\hat{S}_{\textbf{3}}=
\sigma^{x}_{1}\sigma^{x}_{2}\sigma^{x}_{4}\sigma^{x}_{3},\nonumber\\
\hat{S}_{\textbf{3}}\hat{S}_{\textbf{1}}&=&
\sigma^{y}_{1}\sigma^{y}_{2}\sigma^{y}_{3} \sigma^{y}_{4}.\;\;
\end{eqnarray}
The plaquette operators keep an arbitrary ground state vector within
ground state. For example, the operation of plaquette operator
$\hat{S}_{\textbf{1}}$ on the vector ground state,
$\Psi_{[g_{1234}]}=[\;|g\rangle_{1},\;|g\rangle_{2}\;,|g\rangle_{3}\;,|g\rangle_{4}\;]^{T},$
gives a matrix,
\begin{eqnarray}
\hat{S}_{\textbf{1}}\Psi_{[g_{1234}]}=\left[
  \begin{array}{cccc}
0 & 0 & i & 0  \\
0 & 0 & 0 & -i  \\
-i & 0 & 0 & 0  \\
0 & i & 0 & 0\\
  \end{array}
\right]\left[
\begin{array}{c}
|g\rangle_{1} \\
|g\rangle_{2} \\
|g\rangle_{3} \\
|g\rangle_{4} \\
\end{array}
\right].
\end{eqnarray}
The operation of concurrence operator reads,
$\hat{\tau}=\hat{S}_{\textbf{3}}\hat{S}_{\textbf{1}}$,
\begin{eqnarray}
\hat{\tau}\Psi_{[g_{1234}]}=\left[
  \begin{array}{cccc}
-1 & 0 & 0 & 0  \\
0 & 1 & 0 & 0  \\
0 & 0 & -1 & 0  \\
0 & 0 & 0 & 1\\
  \end{array}
\right]\left[
\begin{array}{c}
|g\rangle_{1} \\
|g\rangle_{2} \\
|g\rangle_{3} \\
|g\rangle_{4} \\
\end{array}
\right].
\end{eqnarray}
The four degenerated ground state are entangled states of four
qubits. Since the concurrence operator now happened to be a physical
observable, quantum entanglement maybe can be directly read out in
experiment.

The textbook definition on quantum statistics of two
indistinguishable particles starts from swapping the positions of
the two particles,
$\hat{P}\psi(r_{1},r_{2})=e^{i\theta}\psi(r_{2},r_{1})$, where
$\theta$ is called a statistical angle and $\hat{P}$ is the exchange
operator. $\theta=2\pi$ defines Boson, $\theta=\pi$ defines Fermion,
while [$\theta=\alpha\pi$,$1<\alpha<2$] defines anyon.

Kitaev honeycomb model Hamiltonian can be equivalently mapped into a
p-wave pairing Hamiltonian \cite{kitaev}\cite{feng}\cite{Nayak}.
This triangular star model can also transform into a fermion pairing
model by inverse Jordan-Wigner transformation. Like the Cooper pair
in BCS model, two spins in this model behaves as typical composite
boson as that defined in Ref.\cite{sylee}. While we define the
composite Boson by spin operators,
\begin{eqnarray}
d_{1}=\hat{\sigma}^{z}_{i}\hat{\sigma}^{z}_{j},\;\;\;d_{2}=\hat{\sigma}^{z}_{k}\hat{\sigma}^{z}_{l},\;\;\;(i\neq
j\neq k \neq l).
\end{eqnarray}
These two composite Boson fulfills the Bosonic commutator $[d_{1},
d_{2}]=0$. The spin operators in the composite Boson can also mapped
into a string operator of fermion by Jordan-Wigner transformation
\cite{feng}. That kind of composite particle only exist in quantum
many body system on lattice. Here we shall put the triangular star
model equivalently on a one dimensional small lattice.

As all know, switching two fermions would generate a negative sign
in front of the collective wave function. A composite particle
composed of two fermions has complicate behavior\cite{sylee}. Here
the composite particle of two spins also demonstrate complicate
statistical behavior beyond Boson. We define a permutation operator
$P[{12;34}]$,
\begin{eqnarray}
P[{12;34}]\Psi(1234)=\Psi(3412)=\eta\Psi(1234).\;\;
\end{eqnarray}
Conventionally if $\eta=+1$, we call the composite particle as a
Boson. The Composite particle is fermion if $\eta=-1$. If $\eta$
bear more complex structure, we might call the composite particle as
exotic composite particles. These permutation operator plays a
similar role as the braiding operator in fraction quantum Hall
system \cite{Nayak}. In the quantum Hall system, the wave function
is Laughlin wave function, while the labels, "1,2,3,4", denote the
position of quasiparticles or quasi-holes. Applying this permutation
operator on the well-known Greenberger-Horne- Zeilinger state
\cite{GHZstate} shows it is symmetric state of double qubit,
\begin{eqnarray}
P[{12;34}]|GHZ\rangle=|\Uparrow\Uparrow\rangle+|\Downarrow\Downarrow\rangle=|GHZ\rangle.
\end{eqnarray}
While the generalized W-state of four qubits \cite{Wstate} is also
symmetric state of double qubit,
\begin{eqnarray}
P[{12;34}]|W\rangle=|\Uparrow\circlearrowright\rangle+|\Uparrow\circlearrowleft\rangle
+|\circlearrowright\Uparrow\rangle+|\circlearrowleft\Uparrow\rangle=|W\rangle.
\end{eqnarray}
But Yeo-Chua's genuine four-qubit entangled state \cite{yeo} has
complicate behavior under the operation of this permutation
operator. We decompose Yeo-Chua's state as the sum of two parts,
$|\chi^{00}\rangle=|\chi_{1}^{00}\rangle+|\chi_{2}^{00}\rangle$,
\begin{eqnarray}
|\chi_{1}^{00}\rangle&=&|\Downarrow\Downarrow\rangle+|\Uparrow\Uparrow\rangle
+|\circlearrowright\circlearrowright\rangle-|\circlearrowleft\circlearrowleft\rangle
+|\circlearrowleft\circlearrowright\rangle+|\circlearrowright\circlearrowleft\rangle,\nonumber \\
|\chi_{2}^{00}\rangle&=&|\Uparrow\Downarrow\rangle-|\Downarrow\Uparrow\rangle.
\end{eqnarray}
The first part is symmetric of double qubit cluster, while the
second part is anti-symmetric state of double qubit cluster,
\begin{eqnarray}
P[{12;34}]|\chi_{1}^{00}\rangle=|\chi_{1}^{00}\rangle,\;\;\;
P[{12;34}]|\chi_{2}^{00}\rangle=-|\chi_{2}^{00}\rangle.
\end{eqnarray}
The permutation operator reveals some fine property of entangled
many qubit state. In this triangular star model, even if the four
states bear the same eigenenergy, their behavior under the
permutation operator is completely different.

The eigenstates of this model was expanded in the Hilbert space of
the product operator of composite particle,
$d_{1}d_{2}={\sigma}^{z}_{i}{\sigma}^{z}_{j}{\sigma}^{z}_{k}{\sigma}^{z}_{l}$.
The four degenerated ground states include one femionic cluster
state and one bosonic cluster states,
\begin{eqnarray}
P[{12;34}]|g\rangle_{2}=(-1)|g\rangle_{2},\;\;
P[{12;34}]|g\rangle_{4}=(+1)|g\rangle_{4}.
\end{eqnarray}
The rest two ground states can map into each other by the
permutation operator,
\begin{eqnarray}
P[{12;34}]|g\rangle_{1}=|g\rangle_{3},\;\;\;\;\;P[{12;34}]|g\rangle_{3}=|g\rangle_{1}.
\end{eqnarray}
So we define a four dimensional vector of ground cluster state,
\begin{eqnarray}
\Psi_{[g_{(1234)}]}=[\;|g\rangle_{1},\;|g\rangle_{2}\;,|g\rangle_{3}\;,|g\rangle_{4}\;]^{T},
\end{eqnarray}
then the permutation operates on the vector states,
\begin{eqnarray}
P[{12;34}]\Psi_{[g_{(1234)}]}=\left[
  \begin{array}{cccc}
0 & 0 & 1 & 0  \\
0 & -1 & 0 & 0  \\
1 & 0 & 0 & 0  \\
0 & 0 & 0 & 1 \\
  \end{array}
\right]\left[
\begin{array}{c}
|g\rangle_{1} \\
|g\rangle_{2} \\
|g\rangle_{3} \\
|g\rangle_{4} \\
\end{array}
\right].
\end{eqnarray}
If we select two states out of the four ground states as basis for
quantum computation, the statistical factor of
$\Psi_{[g_{1},g_{3}]}=[\;|g\rangle_{1},\;|g\rangle_{3}\;]^{T}$ is a
Pauli matrix
\begin{eqnarray}
\eta^{g13}_{_{[{12;34}]}}=\sigma^{x}.
\end{eqnarray}
For the other vector wave function of
$\Psi_{[g_{2},g_{4}]}=[\;|g\rangle_{2},\;|g\rangle_{4}\;]^{T},$ the
statistical factor of $\Psi_{[g_{2},g_{4}]}$ is another Pauli matrix
\begin{eqnarray}
\eta^{g24}_{_{[{12;34}]}}=\sigma^{z}.
\end{eqnarray}
The statistical property of the nearest excited state above zero
energy is similar to ground state,
\begin{eqnarray}
P[{12;34}]|e\rangle_{11}&=&|e\rangle_{11},\;\;\;P[{12;34}]|e\rangle_{14}=-|e\rangle_{14}.\nonumber\\
P[{12;34}]|e\rangle_{12}&=&|e\rangle_{13},\;\;\;P[{12;34}]|e\rangle_{13}=|e\rangle_{12}.
\end{eqnarray}
The four degenerate zero energy states and the highest excited
states are all bosonic cluster states,
\begin{eqnarray}
P[{12;34}]|o\rangle_{i}&=&|o\rangle_{i},\;\;i=1,2,3,4;\nonumber\\
P[{12;34}]|e\rangle_{j}&=&|e\rangle_{j},\;\;j=15,16.
\end{eqnarray}
Their corresponding statistical factor are
\begin{eqnarray}
\eta^{o}_{_{[{12;34}]}}=1,\;\;\;\;\;\eta^{15}_{_{[{12;34}]}}=1.
\end{eqnarray}
The first excited state has two degenerated states. Permutating two
clusters produce a statistical factor beyond boson and fermions,
\begin{eqnarray}
P[{12;34}]|e\rangle_{9}&=&e^{i\pi}[\sigma^{x}_{3}\sigma^{x}_{4}]|e\rangle_{10},\;\;\nonumber\\
P[{12;34}]|e\rangle_{10}&=&e^{i\pi}[\sigma^{x}_{3}\sigma^{x}_{4}]|e\rangle_{9},
\end{eqnarray}
For the vector wave function of the first excited states,
$\Psi_{[e_{9},e_{10}]}=[\;|e\rangle_{9},\;|e\rangle_{10}\;]^{T},$
the statistical factor of double spin clusters is
\begin{eqnarray}
\eta^{e9}_{_{[{12;34}]}}=e^{i\pi}\sigma^{x}_{3}\sigma^{x}_{4}\left[
\begin{array}{cc}
0 & 1\\
1 & 0 \\
\end{array}
\right].
\end{eqnarray}
As all know, logic gate operator in quantum computation can be
expanded by Pauli spin matrices. Here the permutation operator of
composite double spins provides one way of implementing quantum
logic gate.

\section{The exotic quantum statistics of frustrated plaquette quasiparticle}

Kitaev honeycomb model was first solved by representing a spin
operator by two Majorana fermions \cite{kitaev}. Later on, a further
study was carried out by Jordan-Wigner transformation \cite{feng}.
Here we use the inverse representation of Jordan-Wigner
transformation to formulate the Majorana fermion operators by spin
operators \cite{yu}. This inverse Jordan-Wigner transformation
fulfills the commutator of fermions as that in the generalized
Jordan-Wigner transformation \cite{Batista}. Here we made a further
mapping from fermion into Majorana fermion.

\begin{figure}
\begin{center}
\includegraphics[width=0.45\textwidth]{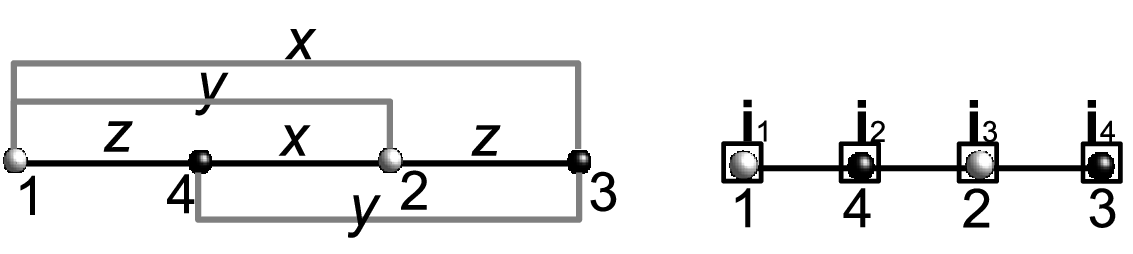}
\caption{\label{tetra4} The one dimensional scheme of the triangular
star model. The initial spatial ordering of the four particle is
[1423]. The coupling bonds connecting the four particles remain the
same as the original triangular star model.}
\end{center}
\vspace{-0.5cm}
\end{figure}

The explicit formulation of inverse Jordan-Wigner transformation
depends on the spatial order of the four particles. We squeeze the
triangular star lattice to a one dimensional chain by keeping the
topology of interacting bonds invariant. The indices
$('1','2','3','4')$ in the Hamiltonian Eq. (\ref{Htri}) are the name
of the four particles instead of its spatial ordering in the one
dimensional chain. We denote the four spatial positions along the
one dimensional chain as
$(\textbf{i}_1<\textbf{i}_2<\textbf{i}_3<\textbf{i}_4)$(Fig.
\ref{tetra4}). If we place particle $'1'$ on $\textbf{i}_4$, the
other three particles would sit on the rest three sites
$(\textbf{i}_1, \textbf{i}_2, \textbf{i}_3)$ before $\textbf{i}_4$.
For a special case of the spatial ordering of the four particles
$[1423]$, i.e.,
$(\textbf{i}_1=1,\;\textbf{i}_2=4,\;\textbf{i}_3=2,\;\textbf{i}_4=3)$,
the inverse Jordan-Wigner transformation defines the spin
representation of eight Majorana fermions,
\begin{eqnarray}\label{1423}
\psi_{1}&=&\sigma_{1}^y,
\;\;\;\;\;\;\;\psi_{4}=\sigma_{4}^x\sigma^z_{1},\;\;\;\;\;\;\;
\psi_{2}=\sigma_{2}^y\sigma^z_{1}\sigma^z_{4},\;\;\; \nonumber\\
b_{1}&=&-\sigma_{1}^x,\;\;\;\;\;
b_{4}=-\sigma_{4}^y\sigma^z_{1},\;\;\;\;\;
b_{2}=-\sigma_{2}^x\sigma^z_{1}\sigma^z_{4},\nonumber\\
\psi_{3}&=&\sigma_{3}^x\sigma^z_{1}\sigma^z_{4}\sigma^z_{2},\;\;\;\;\;
b_{3}=-\sigma_{3}^y\sigma^z_{1}\sigma^z_{4}\sigma^z_{2}.
\end{eqnarray}
The string operator of spins for Majorana fermion exactly fulfills
the commutator of the original Majorana fermions,
$b_{i}^{\dag}=b_{i}$, $\psi_{i}^{\dag}=\psi_{i}$, $\{b_{i},
b_{j}\}=\delta_{ij}$, $\{\psi_{i}, \psi_{j}\}=\delta_{ij}$,
$\{b_{i}, \psi_{j}\}=0$. The Hamiltonian Eq. (\ref{Htri}) under the
inverse Jordan-Wigner transformation Eq. (\ref{1423}) has the
following formulation,
\begin{eqnarray}\label{Hfermi}
H&=&iJ_{x} b_{2}b_{4} -iJ_{y}[\hat{S}_{\textbf{2}}B_{23}]b_{3}b_{4}
+iJ_{z}B_{14}b_{1}b_{4} \nonumber\\
&+& iJ_{x}[\hat{S}_{\textbf{1}}\hat{S}_{\textbf{2}}]b_{1}b_{3}
-iJ_{y}[\hat{S}_{\textbf{3}}B_{14}]b_{1}b_{2}
+iJ_{z}B_{23}b_{2}b_{3} \nonumber \\
&+&J_{p}\hat{S}_{\textbf{1}}\hat{S}_{\textbf{2}}
+J_{p}\hat{S}_{\textbf{2}}\hat{S}_{\textbf{3}}
+J_{p}\hat{S}_{\textbf{3}}\hat{S}_{\textbf{1}},
\end{eqnarray}
where $B_{14}=i\psi_{1}\psi_{4}$ and $B_{23}=i\psi_{2}\psi_{3}$ are
the quantum bond operators on the $\sigma^{z}_{i}\sigma_{j}^{z}$
bond. The inverse Jordan-Wigner transformation is based on the
string operator of  $\sigma^{z}_{i}\sigma_{j}^{z}$.
$B_{14}=i\psi_{1}\psi_{4}$ and $B_{23}=i\psi_{2}\psi_{3}$ commute
with Hamiltonian. This can be checked by directly calculating the
commutator term by term. For a more direct understanding, $B_{14}$
and $B_{23}$ are the pair operator of two fermions, it behaves like
a boson as a composite operator. They both are the product of two
$\psi-$type fermions, thus they commute with all $b-$type fermion
operators. $B_{14}$ commutes with $B_{23}$. $B_{14}$ and $B_{23}$
commutes with the product operator of two plaquette operators. This
can be checked by the fermionic representation of the plaquette
operators,
\begin{eqnarray}
\hat{S}_{\textbf{1}}&=&b_{1}\psi_{1}\psi_{2}b_{3},\;\;\;\;
\hat{S}_{\textbf{2}}=\psi_{4}b_{4}b_{2}\psi_{3},\;\;\nonumber\\
\hat{S}_{\textbf{3}}&=&\psi_{1}\psi_{2}b_{2}b_{4},\;\;\;\;
\hat{S}_{\textbf{4}}=b_{1}\psi_{3}b_{3}\psi_{4}.\;\;
\end{eqnarray}
The fourth plaquette operator
$\hat{S}_{\textbf{4}}=\sigma^{y}_{1}\sigma^{z}_{3}\sigma^{x}_{4}$
runs across the outer boundary of the triangular star.
$\hat{S}_{\textbf{4}}$ is equivalent to the product of the other
three plaquette operators,
$\hat{S}_{\textbf{1}}\hat{S}_{\textbf{2}}\hat{S}_{\textbf{3}}=\hat{S}_{\textbf{4}}$.
The product of two plaquette operators equals to the product of two
conserved bond operator $B_{14}$ and $B_{23}$,
\begin{equation}
\hat{S}_{\textbf{1}}\hat{S}_{\textbf{3}}=-B_{14}B_{23},\;\;\;\;\;\;
\hat{S}_{\textbf{2}}\hat{S}_{\textbf{4}}=-B_{14}B_{23}.\;\;
\end{equation}
Every conserve quantum operator can be handled as good quantum
number.

In the Kitaev honeycomb model, the fermionic representation of
Kitaev Hamiltonian is a p-wave pairing Hamiltonian \cite{feng}. The
fermionic representation of the triangular star model is beyond a
p-wave pairing Hamiltonian. We defined four complex fermions from
Majorana fermions,
\begin{eqnarray}\label{c=b+b}
c_{a}^\dag&=&\frac{1}{2}(b_{1}-ib_{3}),\;\;\;\;\;\;
c_{a}=\frac{1}{2}(b_{1}+ib_{3}),\nonumber\\
c_{b}&=&\frac{1}{2}(b_{2}+ib_{4}),\;\;\;\;\;\;
c_{b}^\dag=\frac{1}{2}(b_{2}-ib_{4}).
\end{eqnarray}
Every Majorana fermion can be expressed by the four complex
fermions. Substituting the complex representation of Majorana
fermions into Hamiltonian Eq. (\ref{Hfermi}) gives
\begin{eqnarray}\label{Hfermic}
H&=&2J_{x}\hat{S}_{\textbf{1}}\hat{S}_{\textbf{2}}c^{\dag}_{a}c_{a}+2J_{x}c^{\dag}_{b}c_{b}+\Delta
c_{a}c_{b} -\Delta^{\ast} c^{\dag}_{a}c^{\dag}_{b}-t^{\ast} c_{a}c^{\dag}_{b} \nonumber\\
&+&tc^{\dag}_{a}c_{b}-2J_{x}+J_{p}\hat{S}_{\textbf{1}}\hat{S}_{\textbf{2}}
+J_{p}\hat{S}_{\textbf{2}}\hat{S}_{\textbf{3}}
+J_{p}\hat{S}_{\textbf{3}}\hat{S}_{\textbf{1}},
\end{eqnarray}
where $\Delta$ is the pairing gap function and $t$ is the hopping
functions,
\begin{eqnarray}
\Delta&=&J_{z}(B_{14}-B_{23})+iJ_{y}(S_{2}B_{23}-S_{3}B_{14}),\nonumber\\
t&=&J_{z}(B_{14}+B_{23})-iJ_{y}(S_{2}B_{23}+S_{3}B_{14}).
\end{eqnarray}
This Hamiltonian is equivalent to a conventional pairing Hamiltonian
for superconductivity that describes the generation and annihilation
of a pair of complex fermions. For a partially fixed pattern of
plaquette operator $\hat{S}_{\textbf{1}}\hat{S}_{\textbf{2}}=1$, the
eigenenergy of the excited quasi-particle of complex fermions is
\begin{eqnarray}
E=\pm
\sqrt{4J^2_{x}+\Delta^{\ast}\Delta}\pm\sqrt{tt^{\ast}}+J_{p}\sum_{ij}\hat{S}_{\textbf{i}}\hat{S}_{\textbf{j}}.
\end{eqnarray}
We expressed this eigenenergy into plaquette operators to get clear
vision on how eigenenergy depends on gauge pattern,
\begin{eqnarray}\label{E}
E&=&\pm \sqrt{4J^2_{x}+2J^2_{z}(1+{S}_{\textbf{3}}{S}_{\textbf{1}})+2J^2_{y}(1+{S}_{\textbf{2}}{S}_{\textbf{1}})}\nonumber\\
&\pm&\sqrt{2J^2_{z}(1-{S}_{\textbf{3}}{S}_{\textbf{1}})+2J^2_{y}(1-{S}_{\textbf{2}}{S}_{\textbf{1}})}\nonumber\\
&+&J_{p}{S}_{\textbf{1}}{S}_{\textbf{2}}
+J_{p}{S}_{\textbf{2}}{S}_{\textbf{3}}
+J_{p}{S}_{\textbf{3}}{S}_{\textbf{1}}.
\end{eqnarray}
For another partially fixed pattern of plaquette operators
$\hat{S}_{\textbf{1}}\hat{S}_{\textbf{2}}=-1$, the eigenenergy of
the excited quasi-particle of the complex fermions is $E'=\pm
\sqrt{4J^2_{x}+t^{\ast}t}\pm\sqrt{\Delta\Delta^{\ast}}.$ Its
dependence on the three plaquette operators is equivalent to Eq.
\ref{E} except that gauge pattern is different.

For the homogeneous gauge pattern,
${S}_{\textbf{1}}={S}_{\textbf{2}}={S}_{\textbf{3}}$, the
eigenenergy of complex fermion excitations is
$E=3J_{p}\pm2\sqrt{J^{2}_{x}+J^{2}_{y}+J^{2}_{z}}$. Under the same
parameter setting as last section, $\{J_{p}=2 J_{x}, \;J_{z}=2
J_{x},\;J_{y}=2 J_{x}\},$ the specific eigenenergy of complex
fermion excitation contributes two levels: one level is $E=12J_x$,
the other is $E=0$. $E=0$ is the vacuum state of complex fermion.
$E=12J_x$ is its highest excited level. The eigenstates with respect
to $E=0$ is $|o\rangle_1$ and $|o\rangle_4$.  The eigenstates with
respect to $\omega=12J_x$ is $|e\rangle_{15}$ and $|e\rangle_{16}$.
For the inhomogeneous gauge pattern,
$\hat{S}_{\textbf{1}}\hat{S}_{\textbf{2}}=-1$, the eigenstates with
respect to $E=0$ is $|o\rangle_2$ and $|o\rangle_3$.

The conventional spin 1/2 particles may form a antisymmetric state
$(|\uparrow\downarrow\rangle-|\downarrow\uparrow\rangle)$ and a
symmetric state
$(|\uparrow\uparrow\rangle+|\downarrow\downarrow\rangle)$. Usually
the singlet state leads to lower energy. The triplet leads to higher
energy. A singlet can be transformed into a triplet state by
flipping the second spin or the first spin,
\begin{eqnarray}
[\;\sigma^{x}_{2}|\uparrow\downarrow\rangle-e^{i\pi}\sigma^{x}_{2}|\downarrow\uparrow\rangle\;]\;\;\;=>\;\;
[\;|\uparrow\uparrow\rangle+|\downarrow\downarrow\rangle\;].
\end{eqnarray}
The singlet state and triplet state above is defined for two spins.
While the spin configuration of the triangle star model includes
four spins. A similar transformation rule as that for a conventional
singlet state and a triplet state also exist between the zero energy
state and the highest energy states of this triangular star model.
Here is we need to flip a pair of spins. Based on the spin
configuration of zero energy states and the highest excited states,
one may extract two unit configuration out of the complete states,
\begin{eqnarray}
A&=&|\Uparrow\Uparrow\rangle
+2|\circlearrowright\circlearrowleft\rangle+2|\circlearrowleft\circlearrowright\rangle
+|\Downarrow\Downarrow\rangle,   \nonumber\\
B&=&5|\circlearrowleft\circlearrowleft\rangle+5|\circlearrowright\circlearrowright\rangle.
\end{eqnarray}
Both the zero energy state and the highest excited state are linear
combination of the two units,
\begin{eqnarray}
 |o\rangle_1=A-B,\;\;\;\;
|e\rangle_{16}=[\sigma^{x}_{3}\sigma^{x}_{4}]A-e^{i\pi}[\sigma^{x}_{3}\sigma^{x}_{4}]B.
\end{eqnarray}
The same algebra relation also exist between the state $|o\rangle_4$
and $|e\rangle_{15}$. The two unit spin configurations of
$|o\rangle_4$ and $|e\rangle_{15}$ are different from that of
$|o\rangle_1$ and $|e\rangle_{16}$. Both $|o\rangle_1$ and
$|e\rangle_{16}$ are bosonic state of double spin clusters. Thus the
zero energy state can be viewed as the generalized singlet state of
double spins, while the highest excited state is the generalized
triplet state of double spins.

\begin{figure}
\begin{center}
\includegraphics[width=0.45\textwidth]{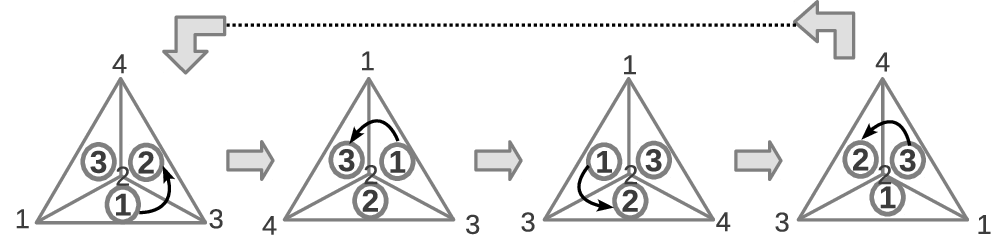}
\caption{\label{loop} The permutating steps for plaquette
quasiparticle ${S}_{\textbf{1}}$ to complete a circle around the
center.}
\end{center}
\vspace{-0.5cm}
\end{figure}

The plaquette quasiparticle coupling terms commute with the other
terms in the Hamiltonian. The ground state is a minimal frustrated
quasiparticle state. $\hat{S}_{\textbf{1}}$ map the four pure ground
states into the same Hilbert space of ground state,
\begin{eqnarray}
\hat{S}_{\textbf{1}}|g\rangle_1&=&i|g\rangle_3,\;\;\;\;\;\;\hat{S}_{\textbf{1}}|g\rangle_2=-i|g\rangle_4,\nonumber\\
\hat{S}_{\textbf{2}}|g\rangle_1&=&-i|g\rangle_3,\;\;\;\;\hat{S}_{\textbf{2}}|g\rangle_2=-i|g\rangle_4.
\end{eqnarray}
The corresponding eigenstates of plaquette operator
$\hat{S}_{\textbf{1}}$ can be constructed by the four ground
eigenstates,
\begin{eqnarray}
|S^+\rangle_A&=&|g\rangle_1+i|g\rangle_3,\;\;\;\;\;\;|S^-\rangle_A=|g\rangle_1-i|g\rangle_3, \nonumber\\
|S^+\rangle_B&=&|g\rangle_2-i|g\rangle_4,\;\;\;\;\;\;|S^-\rangle_B=|g\rangle_2+i|g\rangle_4,
\end{eqnarray}
$|S^{\pm}\rangle_\mu$ are the eigenstates of $\hat{S}_{\textbf{1}}$
and $\hat{S}_{\textbf{2}}$. their corresponding eigenenergy is
$\pm1$, i.e.,
\begin{eqnarray}
\hat{S}_{\textbf{1}}|S^+\rangle_A&=&+1|S^+\rangle_A,\;\;\;\;\;\;\hat{S}_{\textbf{1}}|S^-\rangle_A=-1|S^-\rangle_A, \nonumber\\
\hat{S}_{\textbf{2}}|S^+\rangle_A&=&-1|S^+\rangle_A,\;\;\;\;\;\;\hat{S}_{\textbf{2}}|S^-\rangle_A=+1|S^-\rangle_A.
\end{eqnarray}
The explicit spin configuration of these eigenstate of plaquette
operator are
\begin{eqnarray}
|S^+\rangle_A&=&[|\Downarrow\circlearrowleft\rangle-|\Uparrow\circlearrowright\rangle+|\Uparrow\circlearrowleft\rangle
-|\Downarrow\circlearrowright\rangle] \nonumber\\
&+&i[|\circlearrowleft\Uparrow\rangle-|\circlearrowright\Downarrow\rangle
+|\circlearrowleft\Downarrow\rangle-|\circlearrowright\Uparrow\rangle],\nonumber\\
|S^+\rangle_B&=&[|\Downarrow\Uparrow\rangle-|\Uparrow\Downarrow\rangle-i|\circlearrowleft\circlearrowleft\rangle
+i|\circlearrowright\circlearrowright\rangle].
\end{eqnarray}
The three plaquette operators define three quasiparticles. These
quasiparticles are eigen-excitation of this quantum spin model. We
define the similar permutation operator to the double spin clusters
to investigate the quantum statistics of the triple spin clusters,
\begin{eqnarray}
P[{{S}_{\textbf{1}};{S}_{\textbf{2}}}]\;\Psi(1234)=
\;P[{1;4}]\Psi(1234)=\eta_{_{[{S}_{\textbf{1}};{S}_{\textbf{2}}]}}\Psi(1234).
\end{eqnarray}
Here $(1234)$ indicates the index of the four spins. The four spins
are indistinguishable particles, so does the plaquette
quasiparticles. If we define the index of the four spins in a
different order, the permutation matrix for exchanging two spins in
the new space order are exactly the same as before. So the index
here is simply one representation, it does not make any differences
on the physics. According to this definition, the permutation
operation on the eigenstates of the plaquette operator is
\begin{eqnarray}
P[{{S}_{\textbf{1}};{S}_{\textbf{2}}}]\;|S^+\rangle_B=
\;P[{1;4}]|S^+\rangle_B=\eta_{_{[{S}_{\textbf{1}};{S}_{\textbf{2}}]}}|S^+\rangle_B.
\end{eqnarray}
The spin configuration after the permutation is
\begin{eqnarray}
P[{1;4}]\;|S^+\rangle_B=
\;[|\circlearrowright\circlearrowright\rangle-|\circlearrowleft\circlearrowleft\rangle-i|\Uparrow\Downarrow\rangle
+i|\Downarrow\Uparrow\rangle].
\end{eqnarray}
Comparing the spin configuration after the permutation with the spin
configuration before the permutation,
\begin{eqnarray}
P[{{S}_{\textbf{1}};{S}_{\textbf{2}}}]\;|S^+\rangle_B=i\sigma^{z}_{1}\sigma^{z}_{2}|S^+\rangle_B.
\end{eqnarray}
we read out the statistical matrix of two quasiparticles
$[{S}_{\textbf{1}};{S}_{\textbf{2}}]$ in ground state
$|S^+\rangle_B$,
\begin{eqnarray}
\eta^{B+}_{_{[{S}_{\textbf{1}};{S}_{\textbf{2}}]}}
=i\sigma^{z}_{1}\sigma^{z}_{2}.
\end{eqnarray}
Since the eigenvalue of ${S}_{\textbf{1}}$ and ${S}_{\textbf{2}}$
are both $+1$ for $|S^+\rangle_B$, the statistical factor has a
simpler formulation. The statistical matrix of two quasiparticles is
different for different eigenstate even if they are in the same
Hilbert space of ground state with the same eigenenergy. We take
another eigenstate $|S^+\rangle_A$ at ground state as an example.
The eigenvalue of ${S}_{\textbf{1}}$ for $|S^+\rangle_A$ is $+1$,
while The eigenvalue of ${S}_{\textbf{2}}$ for $|S^+\rangle_A$ is
$-1$. In this case, the statistical matrix is much more complex than
$\eta^{B+}_{_{[{S}_{\textbf{1}};{S}_{\textbf{2}}]}}$.
$|S^+\rangle_A$ can be represented by a matrix state,
\begin{eqnarray}
|S^+\rangle_A=\left(
  \begin{array}{cccccccc}
1 & 0 & 0 & 0 & 0 & 0 & 0 & 0 \\
0 & -1 & 0 & 0 & 0 & 0 & 0 & 0 \\
0 & 0 & 1 & 0 & 0 & 0 & 0 & 0 \\
0 & 0 & 0 & -1 & 0 & 0 & 0 & 0 \\
0 & 0 & 0 & 0 & i & 0 & 0 & 0 \\
0 & 0 & 0 & 0 & 0 & -i & 0 & 0 \\
0 & 0 & 0 & 0 & 0 & 0 & i & 0 \\
0 & 0 & 0 & 0 & 0 & 0 & 0 & -i \\
  \end{array}
\right)
\left(   \begin{array}{c}
|\Downarrow\circlearrowleft\rangle  \\
|\Uparrow\circlearrowright\rangle   \\
|\Uparrow\circlearrowleft\rangle   \\
|\Downarrow\circlearrowright\rangle \\
|\circlearrowleft\Uparrow\rangle   \\
|\circlearrowright\Downarrow\rangle  \\
|\circlearrowleft\Downarrow\rangle  \\
|\circlearrowright\Uparrow\rangle  \\
  \end{array}
\right)\;.
\end{eqnarray}
The explicit spin configuration after permutating the two plaquette
quasi-particles for this state reads
\begin{eqnarray}
&&P[{{S}_{\textbf{1}};{S}_{\textbf{2}}}]|S^+\rangle_A=[|\circlearrowright\Downarrow\rangle-|\circlearrowleft\Uparrow\rangle+|\Uparrow\circlearrowleft\rangle
-|\Downarrow\circlearrowright\rangle] \nonumber\\
&&+i[|\Uparrow\circlearrowright\rangle-|\Downarrow\circlearrowleft\rangle
+|\circlearrowleft\Downarrow\rangle-|\circlearrowright\Uparrow\rangle].
\end{eqnarray}
We also represent this spin configuration
$P[{{S}_{\textbf{1}};{S}_{\textbf{2}}}]|S^+\rangle_A$ by the matrix
formulation,
\begin{eqnarray}
P|S^+\rangle_A=\left(
  \begin{array}{cccccccc}
-i & 0 & 0 & 0 & 0 & 0 & 0 & 0 \\
0 & i & 0 & 0 & 0 & 0 & 0 & 0 \\
0 & 0 & 1 & 0 & 0 & 0 & 0 & 0 \\
0 & 0 & 0 & -1 & 0 & 0 & 0 & 0 \\
0 & 0 & 0 & 0 & -1 & 0 & 0 & 0 \\
0 & 0 & 0 & 0 & 0 & 1 & 0 & 0 \\
0 & 0 & 0 & 0 & 0 & 0 & i & 0 \\
0 & 0 & 0 & 0 & 0 & 0 & 0 & -i \\
  \end{array}
\right) \left(   \begin{array}{c}
|\Downarrow\circlearrowleft\rangle  \\
|\Uparrow\circlearrowright\rangle   \\
|\Uparrow\circlearrowleft\rangle   \\
|\Downarrow\circlearrowright\rangle \\
|\circlearrowleft\Uparrow\rangle   \\
|\circlearrowright\Downarrow\rangle  \\
|\circlearrowleft\Downarrow\rangle  \\
|\circlearrowright\Uparrow\rangle  \\
  \end{array}
\right)\;.
\end{eqnarray}
Performing some matrix algebra for solving the equation,
$P[{{S}_{\textbf{1}};{S}_{\textbf{2}}}]\;|S^+\rangle_A=\eta^{A+}_{_{[{S}_{\textbf{1}};{S}_{\textbf{2}}]}}|S^+\rangle_A$,
we derive the statistical matrix,
\begin{eqnarray}
\eta^{A+}_{_{[{S}_{\textbf{1}};{S}_{\textbf{2}}]}}=\left(
  \begin{array}{cccccccc}
-i & 0 & 0 & 0 & 0 & 0 & 0 & 0 \\
0 & -i & 0 & 0 & 0 & 0 & 0 & 0 \\
0 & 0 & 1 & 0 & 0 & 0 & 0 & 0 \\
0 & 0 & 0 & 1 & 0 & 0 & 0 & 0 \\
0 & 0 & 0 & 0 & i & 0 & 0 & 0 \\
0 & 0 & 0 & 0 & 0 & i & 0 & 0 \\
0 & 0 & 0 & 0 & 0 & 0 & 1 & 0 \\
0 & 0 & 0 & 0 & 0 & 0 & 0 & 1 \\
  \end{array}
\right)\;.
\end{eqnarray}
This is a highly nontrivial statistical matrix. If the statistical
matrix is a positive identity matrix, the two quasiparticles are
boson. If the statistical matrix is a negative identity matrix,  the
two quasiparticles are fermion. While
$\eta^{A+}_{_{[{S}_{\textbf{1}};{S}_{\textbf{2}}]}}$ has complex
elements. Following the same procedure, we can find the statistical
matrix of other plaquette quasiparticles in other eigenstate. If
quasiparticle ${S}_{\textbf{1}}$ complete a closed trajectory and
returns to its starting point (Fig \ref{loop}), then the product of
a series of statistical matrices must fulfill the relationship,
\begin{eqnarray}
P[{{S}_{\textbf{3}};{S}_{\textbf{2}}}]P[{{S}_{\textbf{2}};{S}_{\textbf{1}}}]
P[{{S}_{\textbf{1}};{S}_{\textbf{3}}}]P[{{S}_{\textbf{1}};{S}_{\textbf{2}}}]=I,
\end{eqnarray}
where $I$ is an identity matrix. This nontrivial statistical
behavior of plaquette quasiparticle has potential application in
constructing non-trivial quantum logical gate.

The three spins in one triangular plaquette are highly entangled
with each other. The entropy of reduced density operator of the
triple spin quasiparticle can be used as quantification of quantum
entanglement. We compute the reduced density operator of
$S^+\rangle_B$ for plaquette operator
$S_{\textbf{2}}=\sigma^{y}_{2}\sigma^{x}_{3}\sigma^{z}_{4}$. The
density operator is calculated by tracing out spin "1", i.e.,
$\rho_{_{S_{\textbf{2}}}}(S^+\rangle_B)=Tr_{_{[1]}}\left
[|S^+\rangle_{BB}\langle{S^+}|]_{[1234]}\right]$,
\begin{eqnarray}
\rho_{_{S_{\textbf{2}}}}(|S^+\rangle_B)=\left(
  \begin{array}{cccc}
1 & i & 0 & 0 \\
-i & -1 & 0 & 0  \\
0 & 0 & 1 & -i  \\
0 & 0 & -i & -1  \\
  \end{array}
\right)\;.
\end{eqnarray}
The eigenvalue of this density operators is
$\vec{\lambda}=[\sqrt{2},\sqrt{2},0,0].$ Only the two non-zero
eigenvalues, $\lambda_{1}=\sqrt{2}$ and $\lambda_{2}=\sqrt{2}$,
contribute to von Neumann entropy,
\begin{eqnarray}
S(\rho)=-\lambda_{1}Log[\lambda_{1}]-\lambda_{2}Log[\lambda_{2}]=-\sqrt{2}Log[{{2}}].
\end{eqnarray}
the numerical value of von Neumann entropy $S(\rho)=-0.980258$. The
quantum entanglement of the three spin in plaquette $S_{\textbf{2}}$
is not trivial. We can continue to calculate the reduced density
operator and von Neumann entropy for other plaquette operators,
$S_{\textbf{3}}=\sigma^{y}_{4}\sigma^{x}_{1}\sigma^{z}_{2}$ and
$S_{\textbf{1}}$.

The reduced density operator of is originally based on the spatial
order $[1423]$. If we modify the spatial ordering of the four spins
in the eigenstate, the final density density operator are exactly
the same as the original spatial ordering. Thus the fours are
essentially not spatially indistinguishable.

\section{Summary}

Composite particle of multi-spin clusters in strongly correlated
quantum many body system bear nontrivial statistics beyond boson and
fermion. We extended the textbook definition for quantum statistics
of two elementary particle to composite particle consist of many
particles. The generalized definition is applied to study the
quantum statistics of double-spin clusters and triple-spin
quasiparticles in a triangular star quantum model. The eigenstates
of this model are genuine entangled four qubit states. This
Hamiltonian spontaneously generated eigenstates which shows some
similar but different internal structure as
Greenberger-Horne-Zeilinger state and Yeo-Chua's genuine entangled
states.

With the complete eigenenergy levels and explicit spin configuration
of eigenstates, we found the zero energy state is an anti-symmetric
state of two double-spin clusters, the highest excited state is a
symmetric state of two double-spin clusters. In zero energy level
and the highest energy level, two double-spin clusters behaves as
boson. While in ground state and other intermediate excited state,
the double-spin cluster shows fermionic behavior. This is partly
because the ground state here is minimally frustrated states. If we
choose different basis vector out of the four degenerated ground
states, the statistical matrix produce the well known Pauli
matrices. While Pauli matrix is convenient for constructing quantum
logic gates.

We introduced frustrated coupling plaquette quasiparticles in this
triangular star model. The triple-spin clusters states are the
eigenstates of this model. Thus the excitation of triple-spin
cluster is well defined quasiparticle. The Hamiltonian is mapped
into a fermion pairing Hamiltonian of complex fermions. The
statistical matrix of two triple-spin clusters on the first excited
state is neither fermion nor boson. This suggest the plaquette
quasiparticle obey exotic quantum statistics. The von Neumann
entropy of the reduced density operator of the triangular plaquette
excitation is also highly nontrivial. The plaquette quasiparticle in
this triangular star model follows the same quantum symmetry as that
in Kitaev honeycomb model, the main difference is here we introduced
frustrations between quasiparticles and the quantum operator of
these quasiparticle is only a half of the length as the hexagonal
plaquette in Kitaev honeycomb model. Here the generalized definition
of quantum statistics out of the textbook's definition may help us
to understand the non-abelian anyon in Kitaev honeycomb model.

\section{Acknowledgment}

This work is supported by the Fundamental Research Funds for the
Central Universities.

\end{document}